\documentclass[aps,prd,twocolumn,superscriptaddress,showpacs,preprintnumbers,amsmath,amssymb]{revtex4-2}
\usepackage[utf8]{inputenc}
\usepackage[dvipsnames]{xcolor}
\usepackage[normalem]{ulem}
\usepackage{comment}
\usepackage{overpic}
\usepackage{multirow}
\usepackage{booktabs} 
\usepackage{amsmath}
\usepackage[colorlinks,linkcolor=red,anchorcolor=blue,citecolor=blue,urlcolor=blue,breaklinks=true]{hyperref}
\usepackage{cleveref}
\newsavebox{\tablebox}
\usepackage{graphicx}
\usepackage{dcolumn}
\usepackage{bm}
\usepackage{gensymb}
\usepackage{tabularx}
\usepackage{booktabs}
\usepackage{multirow}
\usepackage{lipsum}
\usepackage{subfigure}
\usepackage{float}
 \newcommand{\BESIIIorcid}[1]{\href{https://orcid.org/#1}{\hspace*{0.1em}\raisebox{-0.45ex}{\includegraphics[width=1em]{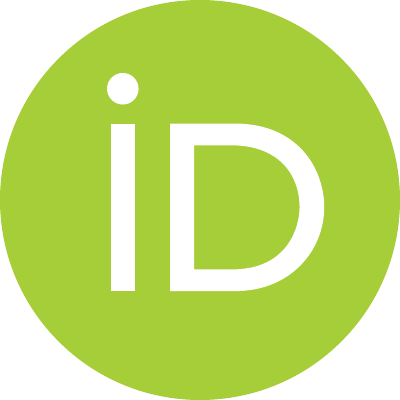}}}}

\begin{document}

\title{\boldmath  Observation of a Doubly-strange Hyperon $\Xi(1720)$ in  $J/\psi\rightarrow{}K^{-}\Sigma^0\bar{\Xi}^{+}+c.c.$}

\author{
M.~Ablikim$^{1}$\BESIIIorcid{0000-0002-3935-619X},
M.~N.~Achasov$^{4,d}$\BESIIIorcid{0000-0002-9400-8622},
P.~Adlarson$^{81}$\BESIIIorcid{0000-0001-6280-3851},
X.~C.~Ai$^{87}$\BESIIIorcid{0000-0003-3856-2415},
C.~S.~Akondi$^{31A,31B}$\BESIIIorcid{0000-0001-6303-5217},
R.~Aliberti$^{39}$\BESIIIorcid{0000-0003-3500-4012},
A.~Amoroso$^{80A,80C}$\BESIIIorcid{0000-0002-3095-8610},
Q.~An$^{77,64,\dagger}$,
Y.~H.~An$^{87}$\BESIIIorcid{0009-0008-3419-0849},
Y.~Bai$^{62}$\BESIIIorcid{0000-0001-6593-5665}
% ...
\\
\vspace{0.2cm}
(BESIII Collaboration)\\
\vspace{0.2cm} {\it
$^{1}$ Institute of High Energy Physics, Beijing 100049, People's Republic of China\\
% ...
$^{4}$ Budker Institute of Nuclear Physics SB RAS (BINP), Novosibirsk 630090, Russia\\
% ...
$^{31}$ INFN Sezione di Ferrara, (A)INFN Sezione di Ferrara, I-44122, Ferrara, Italy; (B)University of Ferrara, I-44122, Ferrara, Italy\\
% ...
$^{39}$ Johannes Gutenberg University of Mainz, Johann-Joachim-Becher-Weg 45, D-55099 Mainz, Germany\\
% ...
$^{62}$ Southeast University, Nanjing 211100, People's Republic of China\\
% ...
$^{64}$ State Key Laboratory of Particle Detection and Electronics, Beijing 100049, Hefei 230026, People's Republic of China\\
% ...
$^{77}$ University of Science and Technology of China, Hefei 230026, People's Republic of China\\
% ...
$^{80}$ University of Turin and INFN, (A)University of Turin, I-10125, Turin, Italy; (B)University of Eastern Piedmont, I-15121, Alessandria, Italy; (C)INFN, I-10125, Turin, Italy\\
$^{81}$ Uppsala University, Box 516, SE-75120 Uppsala, Sweden\\
% ...
$^{87}$ Zhengzhou University, Zhengzhou 450001, People's Republic of China\\

\vspace{0.2cm}
$^{\dagger}$ Deceased\\
% ...
$^{d}$ Also at the Novosibirsk State University, Novosibirsk, 630090, Russia\\
% ...
}

}

\begin{abstract}

Based on a sample of $(10087 \pm 44) \times 10^6$ $J/\psi$ events collected with the BESIII detector at BEPCII, we report the first observation of the decay $J/\psi \rightarrow K^- \Sigma^0 \bar{\Xi}^++c.c.$. A partial wave analysis is performed to investigate the involved excited states. In addition to the well-established $\Xi(1690)$, a new doubly-strange hyperon $\Xi(1720) $ is observed decaying to $K^- \Sigma^0$ with a mass of $1721.0 \pm 5.2_{\rm stat.} \pm 3.4_{\rm syst.} ~{\rm MeV}/c^2$ and a width of $31.3 \pm 18.3_{\rm stat.} \pm 15.4_{\rm syst.} ~{\rm MeV}$, with a statistical significance exceeding $10\sigma$.  The spin-parity hypothesis testing across various quantum number configurations reveals that the spin-parity of $\Xi(1720)$ favors $J^P = {\frac{3}{2}}^+$. Furthermore, the branching fraction of $J/\psi \rightarrow K^- \Sigma^0 \bar{\Xi}^++c.c.$ is determined to be $(2.68 \pm 0.04_{\rm stat.} \pm 0.17_{\rm syst.}) \times 10^{-5}$.
\end{abstract}
\pacs{11.80.Et, 13.25.Gv, 14.20.Jn, 13.20.Gd }
\maketitle
The quark model~\cite{Gell-Mann:1964ewy}, an outstanding achievement of the last century, provides a robust framework for describing
the hadron spectrum. However, baryon spectroscopy remains incomplete, 
as many states predicted within the SU(3) multiplets are either undiscovered or not well established, particularly in the case of cascade hyperons with strangeness $S=-2$. 
The study of excited states,  $\Xi^*$, represents a critical area of research in hadron physics, offering profound insights into the non-perturbative aspects of quantum chromodynamics (QCD) and the underlying structure of baryons.

Theoretical models \cite{Capstick:2000qj,Glozman:1997ag,Glozman:1995fu,Menapara:2021dzi} predicted the existence of more than thirty excited  $\Xi$ states, yet only a handful have been experimentally observed and confirmed. This scarcity is largely due to their small production cross sections and the complex topologies of their final states. Historically, most observations and measurements have been derived from bubble chamber experiments or diffractive $K^- p$ interactions. Notably, aside from the $\Xi(1530)$, all excited $\Xi$ resonances are rated below four stars in the Particle Data Group (PDG)~\cite{ParticleDataGroup:2024cfk}, indicating that they have not been well established.

A full picture of the  precision spectroscopy of excited $\Xi$ baryons provides the essential insights into QCD, elucidating confinement mechanisms and exotic multi-quark dynamics.
The high statistics of $J/\psi$ and $\psi(3686)$ events ~\cite{BESIII:2021cxx,BESIII:2017tvm} accumulated at the BESIII experiment   provide exceptional opportunities for studying excited  hyperon states.  Earlier analyses of $\psi(3686) \rightarrow K^- \Lambda \bar{\Xi}^+ $  ~\cite{BESIII:2015dvj,BESIII:2023mlv} events identified the two excited hyperons, $\Xi(1690)$ and $\Xi(1820)$.   
Although the phase space of $J/\psi$ decays is limited, the world's largest sample of  10 billion $J/\psi$ events ~\cite{BESIII:2021cxx} produced from the $e^+e^-$ annihilations   offers a unique chance to  search for the new excited $\Xi$ states, which are essential for advancing our understanding of baryon spectroscopy.

 In this Letter, we  report the first observation of the decay $J/\psi \rightarrow K^- \Sigma^0\bar{\Xi}^+$ using the 10 billion $J/\psi$ events  accumulated  with the BESIII detector, which is described in detail in Ref.~\cite{BESIII:2009fln}.   The comprehensive partial wave analysis (PWA)  finds that  a double strangeness hyperon state  $\Xi(1720)$ is observed in $K^- \Sigma^0$ mass spectrum
 and  its spin-parity  is determined to be  $J^P={\frac{3}{2}}^+$. 
The inclusion of charge conjugate modes is always implied unless explicitly stated otherwise.

 Given the low momenta of the final states resulting from the limited phase space in the decay  $J/\psi \rightarrow K^- \Sigma^0 \bar{\Xi}^+$,  a partial reconstruction method is employed to enhance the event selection efficiency and increase the statistical  power.
 In the decay $J/\psi \rightarrow K^- \Sigma^0 \bar{\Xi}^+$ with $\bar{\Xi}^+\rightarrow \bar{\Lambda}\pi^+,\bar{\Lambda}\rightarrow \bar{p}\pi^+ $ ,  we  reconstruct only  $K^-$ and $\bar{\Xi}^+$. The   $\Sigma^0$ is treated as a missing particle,  inferred from its recoil against $K^-$ and $\bar{\Xi}^+$. 
 Charged tracks detected in the multi-layer drift chamber  (MDC) are required to be within a polar angle ($\theta$)
range of $|\cos\theta|  < $ 0.93, where $\theta$ is defined with respect to the $z$-axis.
The distance of closest approach to the interaction point (IP) must satisfy  $|V_{z} |< 20$ cm along the $z$-axis,
and $V_{xy}< 10$ cm in the transverse plane, because $\bar{\Lambda}$ and $\bar{\Xi}$
are long-lived particles. 
Particle identification (PID) for charged tracks combines measurements of the
energy deposited in the MDC ($\mathrm{d}E/\mathrm{d}x$) and the flight time in the time-of-flight system (TOF) to form
likelihoods $\mathcal{L}(h) (h = p, K, \pi)$ for each hadron $h$ hypothesis.   
A track is identified as $p$ (or $K$,$\pi$) when the corresponding  hypothesis is the largest likelihood.

Candidate events with one $\bar{p}$, one $K^-$ and two $\pi^+$s are retained for further analysis. The $K^-$ is additionally required to satisfy   $V_{xy} <1$ cm and $|V_{z}| <5$ cm as it originates from the IP directly.   To reconstruct the  $\bar{\Lambda}$ candidate,  
$\bar{p}\pi^+$ pairs sharing a common vertex are selected and subjected to a secondary vertex fit.
The $\bar{\Xi}^+$  is reconstructed with the $\bar{\Lambda}$ candidate whose  invariant mass is closest to the nominal $\Lambda$ mass in the PDG~\cite{ParticleDataGroup:2024cfk},
combined with  an additional $\pi^+$ through a secondary vertex fit. 
 The candidate $\bar{\Xi}^+$ is required to have 
 a  decay length greater than 0, and  its mass must be within 10 MeV/$c^2$ of  the nominal $\bar{\Xi}^+$  mass.  
 To further suppress non-peaking background events under the $\bar{\Lambda}$ and $\bar{\Xi}^+$ peaks, the mass difference between the reconstructed $\bar{\Lambda}$ and $\bar{\Xi}^+$ is required to satisfy
  $|(m_{\bar{\Lambda }\pi^+}-m_{\bar{\Xi}^+}) - (m_{\bar{p}\pi^-}-m_{\bar{\Lambda}})|<4$ MeV/$c^2$, where $m_{\bar{\Xi}^+}$ and $m_{\bar{\Lambda}}$ are the nominal masses from  PDG~\cite{ParticleDataGroup:2024cfk}.  Additionally, 
the invariant mass  recoiling against $K^-$, corresponding to the  mass spectrum of $\Sigma^0$ and $\bar{\Xi}^+$,  is required to be less than $2.58 ~\text{GeV}/c^2$ to discard events with low-momentum kaons, which are typically not well reconstructed due to limitations in the detector's resolution and efficiency.

\begin{figure}[htbp]
\centering
\includegraphics[width=0.39\textwidth]{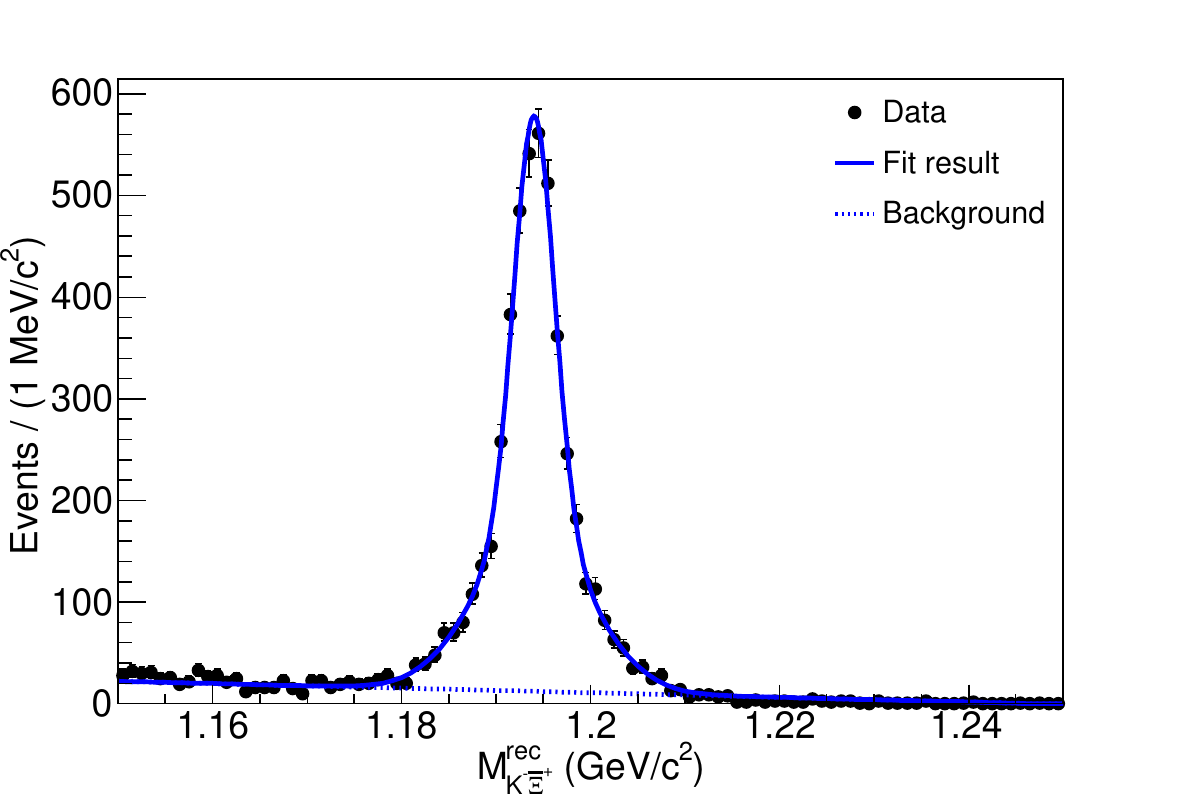}
\caption{The distribution of the invariant mass  recoiling against $K^-\bar{\Xi}^+$, with the fit result superimposed.} 
\label{fitmiss}
\end{figure}

After the above requirements,  a clear  $\Sigma^0$ peak is   observed in  the mass spectrum recoiling against  $K^-\bar{\Xi}^+$ as shown in Fig.~\ref{fitmiss}.  
Potential background events are examined using a Monte Carlo (MC) sample consisting of 10 billion $J/\psi$ inclusive decays, generated with the LUNDCHARM and EVTGEN models~\cite{Chen:2000tv,Lange:2001uf,Ping:2008zz}.
It is found that  there are no peaking background events, which imply that the  sideband events  around the $\Sigma^0$ peak can be reasonably used to estimate the background contributions.   
To extract the number of $\Sigma^0$ events, an unbinned maximum likelihood fit, which yields $4545\pm73$  events, is applied to the  mass spectrum  recoiling against  $K^-\bar{\Xi}^+$ with a double Gaussian function for the signal plus a 
second order Chebychev polynomial as the background function,   as displayed in Fig. ~\ref{fitmiss}.   

\begin{figure}[htbp]
\centering
\includegraphics[width=0.39\textwidth]{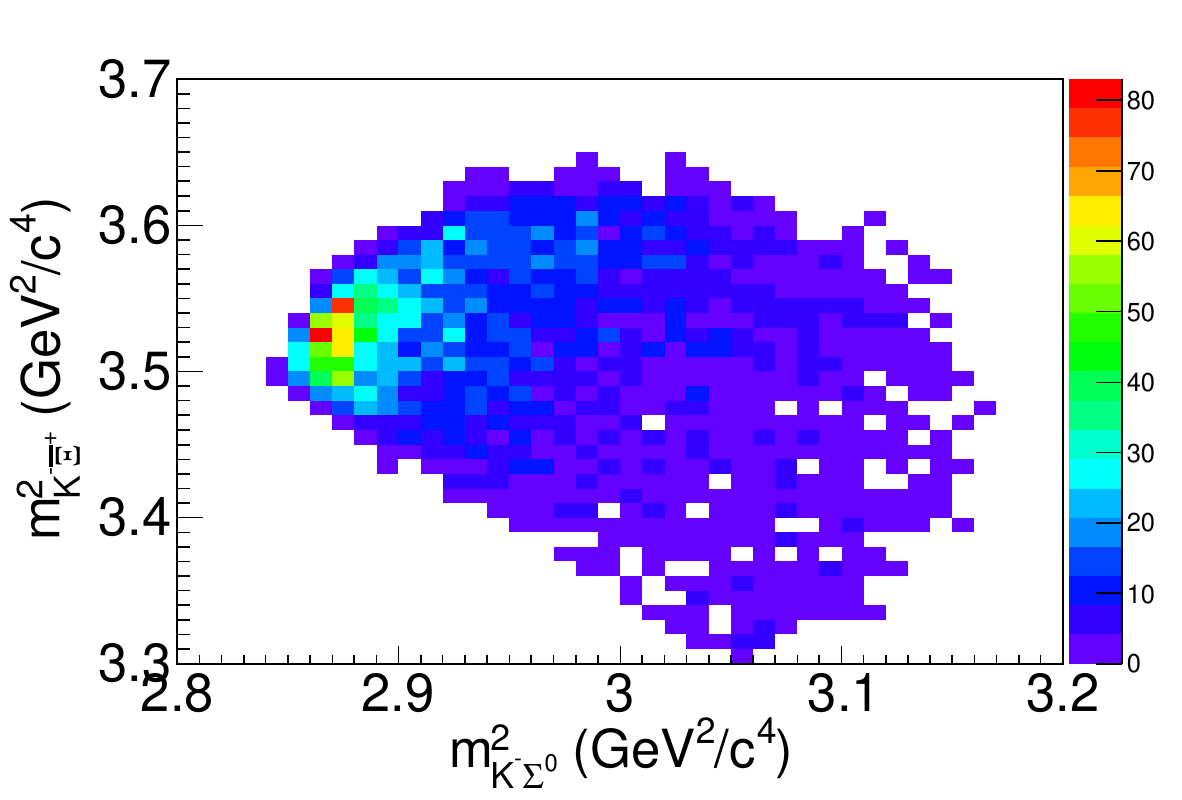}
\caption{The Dalitz plot of $m^2_{K^-\bar{\Xi}^+}$ versus $m^2_{K^-\Sigma^0}$.} 
\label{dalitz}
\end{figure}

\begin{figure*}[!htbp]
\centering
\includegraphics[width=0.33\textwidth]{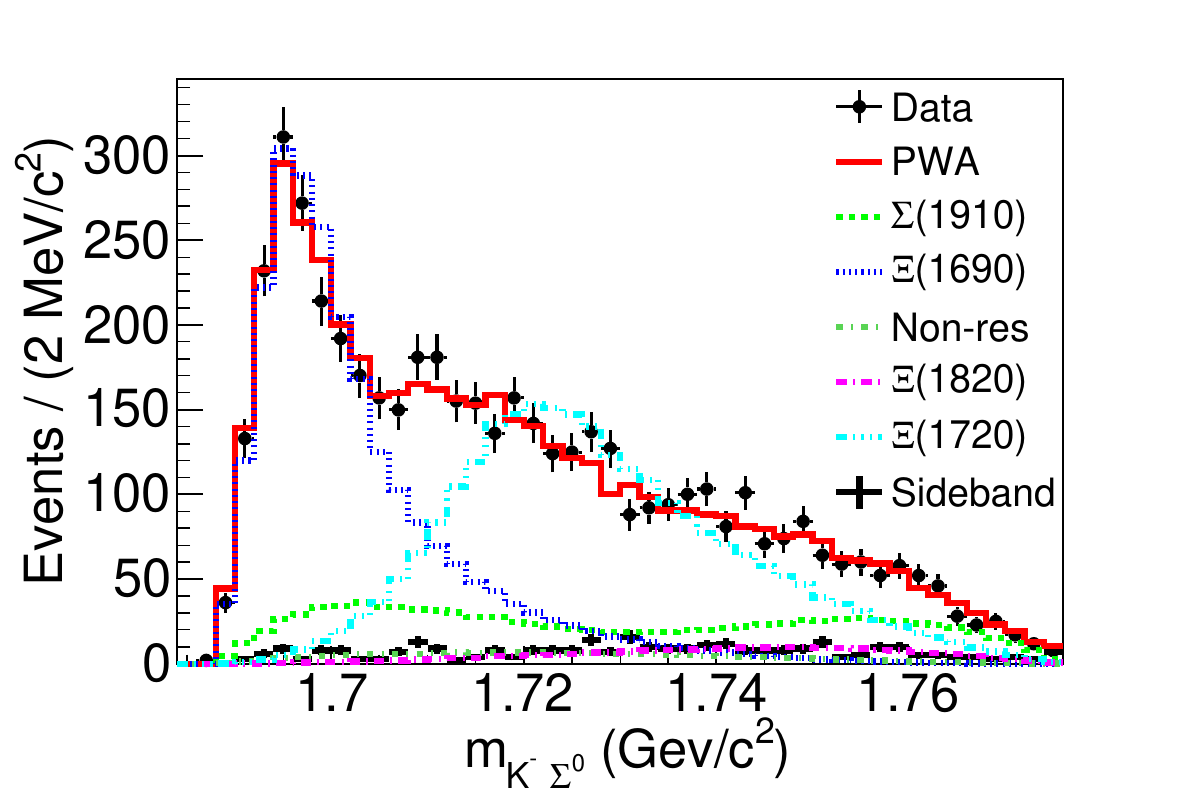}
\put(-140,92){$(a)$}
\includegraphics[width=0.33\textwidth]{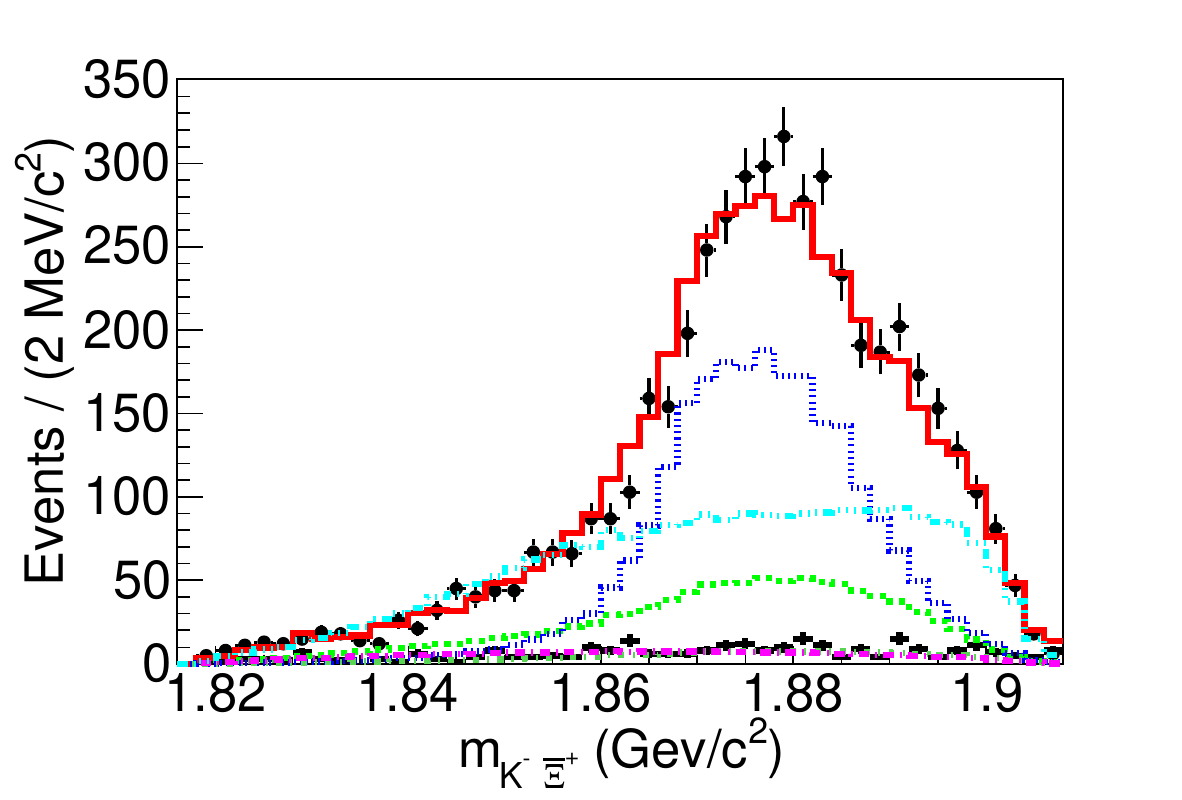}
\put(-140,92){$(b)$}
\includegraphics[width=0.33\textwidth]{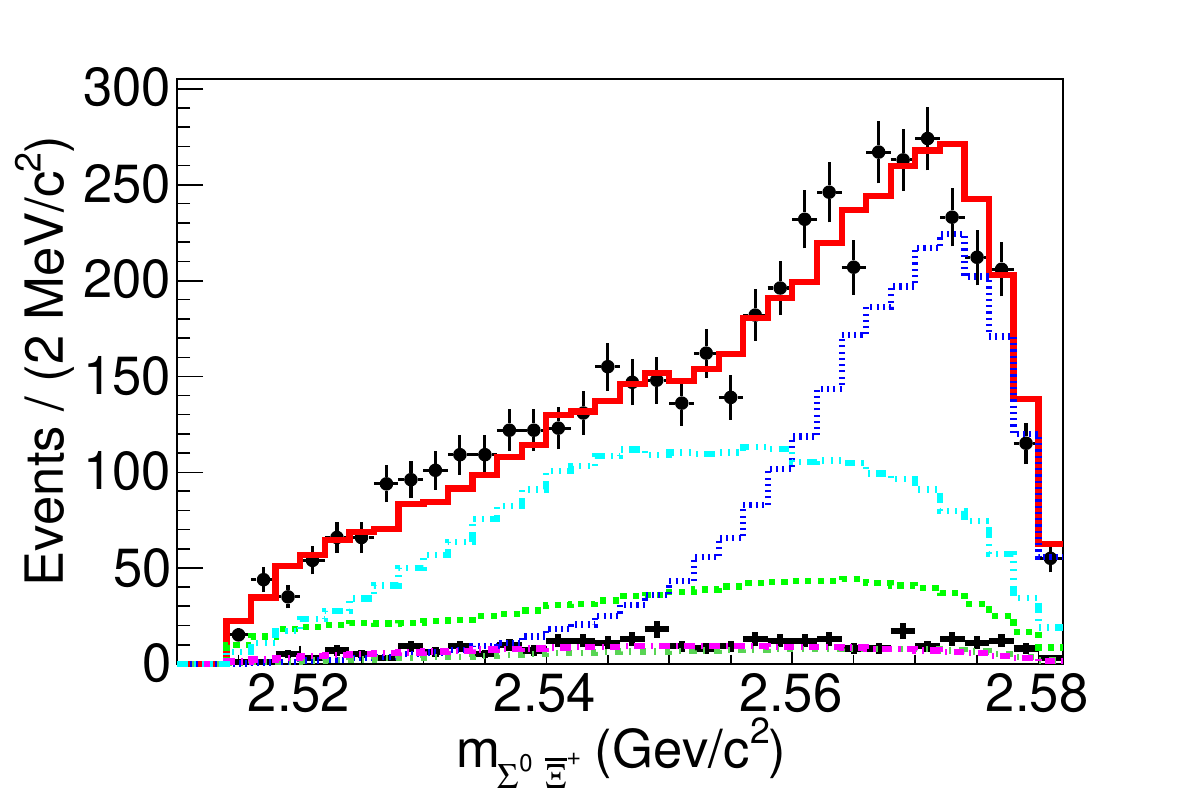}
\put(-140,92){$(c)$}

\includegraphics[width=0.33\textwidth]{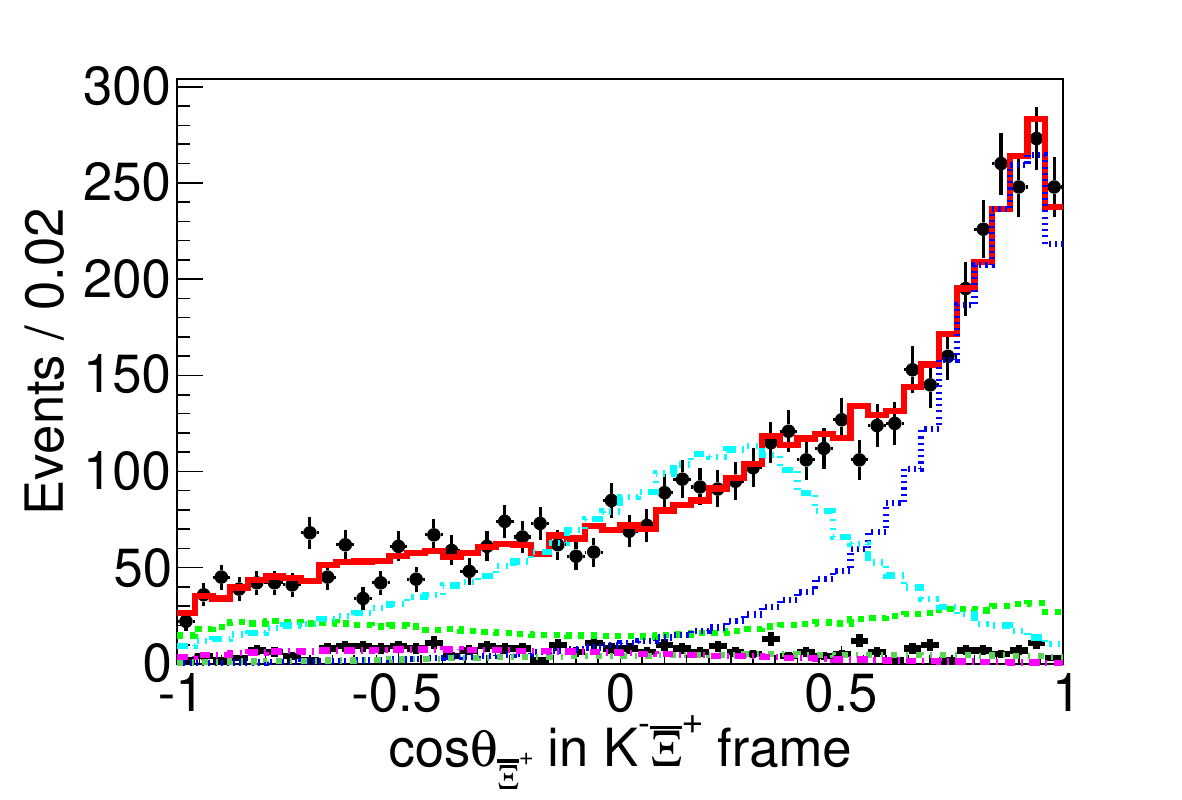}
\put(-140,92){$(d)$}
\includegraphics[width=0.33\textwidth]{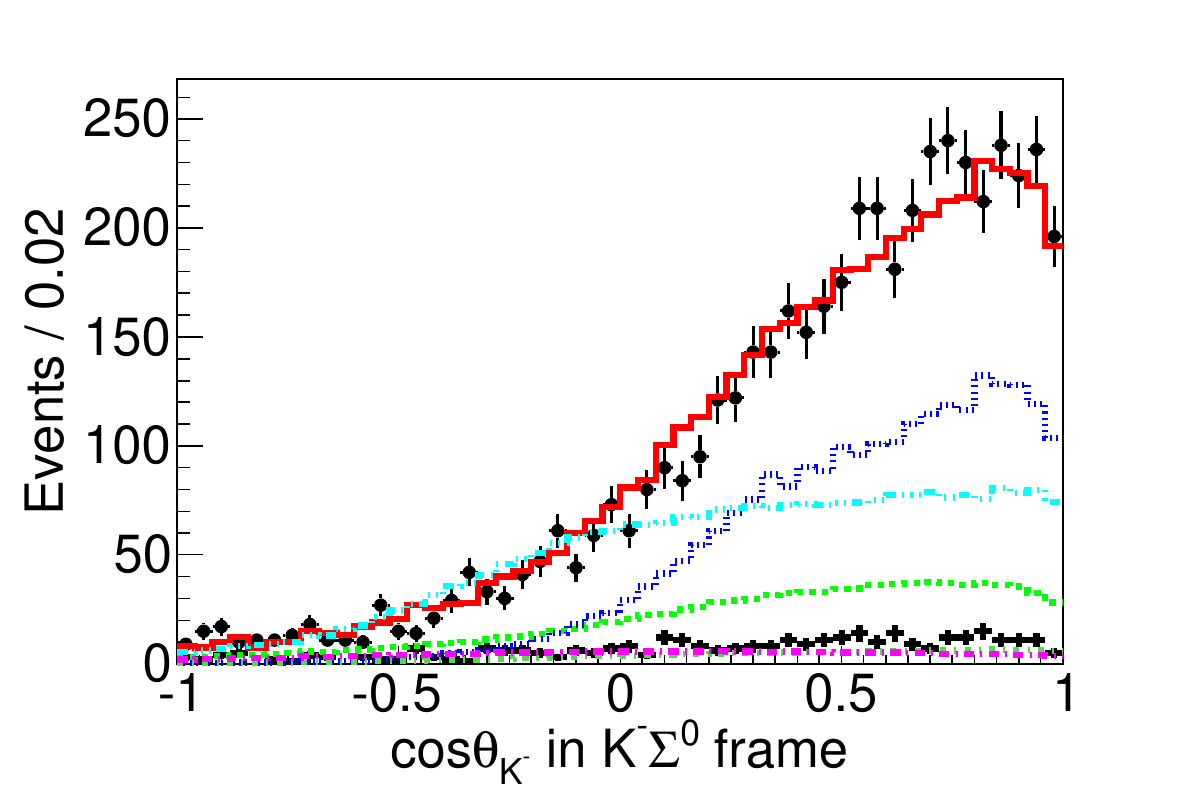}
\put(-140,92){$(e)$}
\includegraphics[width=0.33\textwidth]{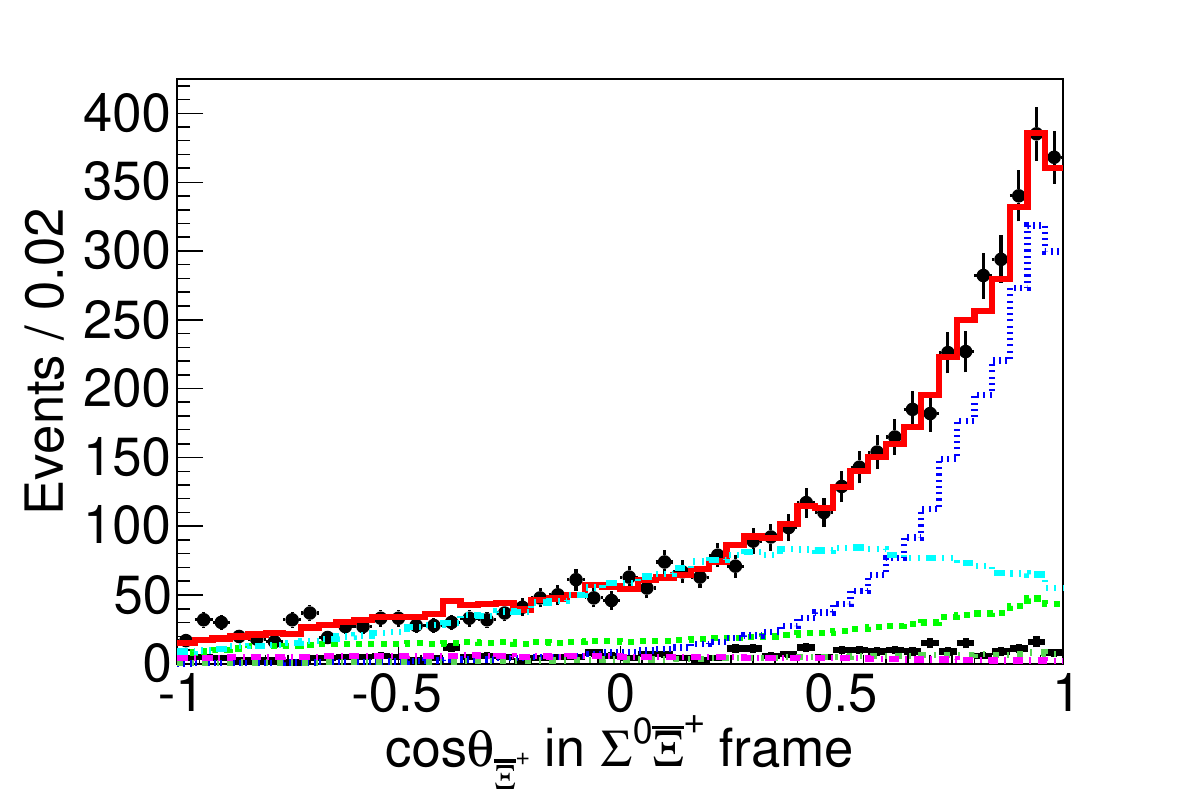}
\put(-140,92){$(f)$}
\caption{ Comparisons between data and PWA fit projections for the invariant mass spectra of $K^-\Sigma^0$ (a), $K^-\bar{\Xi}^+$ (b) and $\Sigma^0\bar{\Xi}^+$ (c), and the angular distributions of 
  $K^-$ in the $K^-\bar{\Xi}^+$  helicity frame (d) ,  $\Sigma^0$ in the $K^-\Sigma^0$ helicity frame (e) and $\Sigma^0$ in the $\Sigma^0\bar{\Xi}^+$  helicity frame (f).  The dots with errors represent for data, the cross with error bars are for background events, the red histograms are for the global fit projections and the other histograms are for the individual contributions from each component. 
} 
\label{PWA}
\end{figure*}

With the requirement of events within the 15 MeV/$c^2$ window around the known mass of $\Sigma^0$~\cite{ParticleDataGroup:2024cfk}, 
a one-constraint (1-C) kinematic fit, constraining the missing mass to  the nominal mass of $\Sigma^0$, is applied to improve mass resolution only. After that the Dalitz plot of the selected 4882 candidate events is presented in Fig.~\ref{dalitz},  where the clear cluster corresponds to
the significant enhancement near the $K^-\Sigma^0$  mass threshold  in Fig.~\ref{PWA} (a), corresponding to the well established $\Xi(1690)$. The number of background events, $335\pm 19$, is  determined  using the sidebands of  $\Sigma^0$, defined as
 1.159  GeV/$c^2$ $<M^{rec}_{K^-\bar{\Xi}^+}<1.174$ GeV/$c^2$ and  1.214  GeV/$c^2$ $<M^{rec}_{K^-\bar{\Xi}^+}<1.229$ GeV/$c^2$. No evident peaking contributions are observed in
the mass spectra of $K^-\Sigma^0$, $K^-\bar{\Xi}^+$ and $\bar{\Xi}^+\Sigma^0$ as shown in Fig.~\ref{PWA}, and  the background level is estimated to be $6.9\%$.

 A PWA of  $J/\psi\rightarrow K^{-}\Sigma^0\bar{\Xi}^{+}$ is  then performed to investigate possible intermediate resonances involved in this process. In the PWA, the amplitudes are constructed using the relativistic covariant tensor amplitude 
formalism ~\cite{Zou:2002ar}, and the maximum likelihood method is used in the PWA with the FDC~\cite{Wang:2004du} package.
The total transition probability, $\omega$, for each event
is obtained from a  linear combination of all partial
wave amplitudes $A_j$ as $\omega = |\Sigma c_jA_j|^2$, where 
the $c_j$ parameters are to be determined by fitting the data.
 The negative log-likelihood function $\ln \mathcal{L}$ can be constructed as 
 \begin{equation}
\ln \mathcal{L} = \sum_{k=1}^{n}{\ln\frac{\omega(\xi_k)}{\int d
\xi\omega(\xi)\epsilon(\xi)} },  
\end{equation}
where $\xi_k$ is the energy-momentum of the final state of
the $k$-th observed event, 
$\epsilon(\xi)$ is the detection efficiency
for the combination $\xi$.
In this analysis, $-\ln\mathcal{L}$ is minimized
to get $c_j$. 
  The final negative log-likelihood value minimized for the data, $S$, is the sum of the log-likelihood values of the events in the $\Sigma^0$ signal region and
background events estimated in the $\Sigma^0$ sideband region with negative weights, defined as, 
\begin{equation}
  S=-\ln\mathcal{L}_{\rm{data}} + \ln\mathcal{L}_{\rm{bg}}.
  \label{eq:likelihood}
\end{equation}

In the PWA, the Breit-Wigner formula of a resonance is written as
$ BW(s)=\frac{1}{M^{2}-s-iM\Gamma}$,
where $M$ is the mass, $\Gamma$ is the constant width, and  $s$ is the invariant mass squared of the decay products of the resonance.

The following form factors, which arise from the internal structure of the resonance as discussed in Refs.~\cite{Penner:2002ma,Liang:2004sd},  are multiplied by the Breit-Wigner function
to account for the  modification of the  Breit-Wigner shape:
\begin{equation}
  J=\frac{1}{2};\quad F_{N}(s)=\frac{\lambda^{4}_{1}}
  {\lambda^{4}_{1}+(s-M^{2})^{2}},
  \label{eqlam1}
\end{equation}

\begin{equation}
  J=\frac{3}{2}, \frac{5}{2};\quad
  F_{N}(s)=e^{\frac{-|s-M^{2}|}
  {\lambda^{2}_{2}}},
  \label{eqlam2}
\end{equation}
where $J$ is the spin, 
and the cut-off parameters $\lambda_{1}$ and $\lambda_{2}$ are set to be 2.0 GeV experimentally.

For the $\Xi^*$, \textit{e.g.} $\Xi(1690)$, located near the mass threshold of $K$ and $\Sigma$, the 
width can be expressed as ~\cite{Vrana:1999nt}:
\begin{equation}
\label{w1690}
\begin{split}
 \Gamma \rightarrow \Gamma \times [(1-f)\times \frac{B^2_L(q_{s,K\Lambda})\rho_{K\Lambda}(s)}{B^2_L(q_{m^2_{\Xi^*},K\Lambda})\rho_{K\Lambda}(m_{\Xi^*}^2)} \\
   + f\times
 \frac{B^2_L(q_{s,K\Sigma})\rho_{K\Sigma}(s)}{B^2_L(q_{m_{\Xi^*}^2,K\Sigma})\rho_{K\Sigma}(m_{\Xi^*}^2)}],
\end{split}
\end{equation}
 where  $f$ represents the fraction of $\Xi^*$ coupling to $K\Sigma$
 and it is optimized for each component in PWA.
 $\rho_{K\Lambda}(s)$ and $\rho_{K\Sigma}(s)$ are the phase space factors for the $K\Lambda$ and  $K\Sigma$ final states, respectively.
$B_L$ is the standard Blatt-Weisskopf barrier factor~\cite{Chung:1993da,Zhu:1999ix} for the decay with orbital angular momentum $L$. 

 All the possible intermediate resonances listed in the PDG~\cite{ParticleDataGroup:2024cfk} have been tested, and only those with a statistical significance exceeding $5\sigma$, $\Xi(1690)$, $\Xi(1820)$ and $\Sigma(1910)$, are included in the PWA. The statistical significance of each resonance is
evaluated by examining the change in the log likelihood
value and the number of free parameters in the fits with and
without the specific resonance.  To account for the contributions from  the phase space,    a non-resonant component is also incorporated.
 However,  the PWA  preliminary result shows a noticeable deviation from the data distribution in the  mass spectrum of $K\Sigma^0$, which suggests that only the known resonant contributions and non-resonant contributions are insufficient to provide a good description of data. 
 
 To address this issue, a new resonant component is introduced to the PWA framework with the hypotheses of various spin-parity($J^P= \frac{1}{2}^{\pm},\frac{3}{2}^{\pm} $). It is found that  $J^P= \frac{3}{2}^{+}$ emerges as the most significant with a statistical significance exceeding 10$\sigma$.  This new resonance is  denoted as $\Xi(1720)$ in the nominal result with an s-dependent width as described in Eq.~(\ref{w1690}) and the resonant parameters are determined to be 
 $m_{\Xi(1720)}=1721.0 \pm 5.2 ~{\rm MeV}/c^2$, and the  width $\Gamma_{\Xi(1720)}= 31.3 \pm 18.3 ~{\rm MeV}$.  The necessity of this resonant contribution could also be seen in the angular comparisons
 between data and PWA fit projections as illustrated in Figs.~\ref{PWA} (d), (e) and (f). 
 The PWA results indicate that $J^P$ of  $\Xi(1690)$ favors 
  $ =\frac{1}{2}^-$ with a mass of $1695.0 \pm 0.9$ ~ MeV/$c^2$, and a width of $12.8 \pm 1.8$~MeV, which are consistent with the previous measurements~\cite{BESIII:2015dvj,BESIII:2023mlv}.  Here  the uncertainties are statistical only. 
  To evaluate the contributions from the  $\Sigma^{*0}$ states around 1.9 GeV/c$^2$ in the $K^-\bar{\Xi}^+$ mass spectrum, the PWA is performed including $\Sigma^{0}(1880)$, $\Sigma^{0}(1910)$ and  $\Sigma^{0}(1915)$. However,   only the contribution from $\Sigma(1910)$ is found to be  significant. Additionally, 
   possible contributions from other $\Xi^*$ resonances are investigated by performing the fits with alternative combinations.  The best fit favors the presence of $\Xi(1820)$ with a statistical significance of 6.9$\sigma$,  providing a better description of data.  In this analysis, the resonance parameters for both $\Xi(1820)$ and $\Sigma(1910)$ are fixed to be the world average values~\cite{ParticleDataGroup:2024cfk}.

  The projections of the data and the baseline fit result onto the mass spectra of $K^-\Sigma^0$, $K^-\bar{\Xi}^+$ and $\bar{\Xi}^+\Sigma^0$, shown in Fig.~\ref{PWA}, indicate that the fit model distribution matches well that of the data.
  With the detection efficiency obtained from the MC simulation in accordance with the PWA results, 
the product branching fraction of $J/\psi\rightarrow K^-\Sigma^0\bar{\Xi}^+$ is calculated with  
  \begin{equation}
\label{eqbr}
\mathcal{B}=\frac{N_{\rm sig}}{N_{J/\psi}\mathcal{B}(\Xi^-\rightarrow \Lambda
\pi^-)\mathcal{B}(\Lambda\rightarrow p \pi^-)\varepsilon },
\end{equation}
where $N_{J/\psi} = (10087\pm44)\times10^6$ is the number of $J/\psi$ events in data~\cite{BESIII:2021cxx}. The branching fractions, $\mathcal{B}(\Xi^-\rightarrow \Lambda
\pi^-)= 0.9989 \pm 0.0004 $ and $\mathcal{B}(\Lambda\rightarrow p \pi^-) = 0.641 \pm 0.005$, are taken from the PDG~\cite{ParticleDataGroup:2024cfk}.
$N_{\rm sig}$   is the number of signal events for  each resonant component, and $\varepsilon$ represents the corresponding detection efficiency.

With the number of  $J/\psi\rightarrow  K^-\Sigma^0\bar{\Xi}^+$ events, $4545\pm73$, from the fit to the mass spectrum recoiling against $K^-\bar{\Xi}^+$ and MC-determined efficiency, 2.86\%, the branching fraction for the decay  $J/\psi\rightarrow  K^-\Sigma^0\bar{\Xi}^+$ is then determined to be  $(2.68\pm0.04\pm0.17 )\times 10^{-5}$.
 In addition, using the fitted signals, $2372\pm684$ and $2542\pm874$, for $\Xi(1690)$ and $\Xi(1720)$, respectively, along with the detection efficiencies,  $3.24\%$ and $1.98\%$, the product branching fractions for these two intermediate states are calculated to be $(1.86\pm0.54\pm0.68)\times 10^{-5}$ and $(1.22\pm0.42\pm0.83)\times 10^{-5}$. Here the first and second uncertainties are statistical and systematic, respectively, and the detection efficiencies used in the branching fraction calculations are obtained from the MC simulations in accordance with the PWA results.

\begin{table}[!htbp]
\begin{center}
\caption{Summary of the systematic uncertainty sources and their corresponding contributions (in \%) to  the branching fractions of  $J/\psi \rightarrow \Xi(1690)^-[\Xi(1720)^-] \bar{\Xi}^+$ and $J/\psi \rightarrow K^- \Sigma^0 \bar{\Xi}^+$.}
\vspace{0.5cm}
\label{sigerr}
\begin{tabular}{l|ccc}\hline\hline
Source & 
$\Xi(1690)^-$& $\Xi(1720)^-$&$K^-\Sigma^0\bar{\Xi}^+$ \\\hline
MDC tracking  & 1.0  &  1.0 & 1.0    \\
PID      & 1.0  &  1.0 & 1.0    \\
$\Xi^-$ reconstruction   & 5.1  &  5.1 & 5.1    \\
Signal fitting    & 3.1& 3.0& 3.1   \\
$\mathcal{B}(\Lambda\rightarrow p \pi^-)$ & 0.8& 0.8& 0.8 \\
Number of $J/\psi$ events  &0.5 &0.5   &0.5 \\
PWA bias  &4.7& 13.4 &- \\
Background  &3.3& 4.9  & -\\
Impact of $\Xi(1820)$  &2.9 &52.8 &-\\
Impact of $\Sigma(1910)$  &13.7 & 20.1&-\\
Extra resonance   &32.3 & 30.3&-\\
Width Para. & 3.3 & 8.9   &-\\
Form factor &   1.2 & 12.3   &-\\\hline
Total         & 36.4& 67.7& 6.2\\\hline\hline

\end{tabular}
\end{center}
\end{table}

\begin{table}[!htbp]
\begin{center}
\caption{Summary of the systematic uncertainty sources and their corresponding contributions to the masses (in unit of MeV/$c^2$) and widths (in unit of MeV) of $\Xi(1690)$ and $\Xi(1720)$, respectively.}
\vspace{0.5cm}
\label{reserr}
\begin{tabular}{l|cccc}\hline\hline
Source & $m_{\Xi(1690)}$& $\Gamma_{\Xi(1690)}$& $m_{\Xi(1720)}$&
$\Gamma_{\Xi(1720)}$\\ \hline

PWA I/O bias &0.1 &0.3  &1.5 & 8.5  \\
Background &0.1 &0.4  &0.1 &0.9   \\
Impacts of $\Xi(1820)$&0.1 & 2.8&1.0 &10.0\\ 
Impacts of $\Sigma(1910)$&0.3& 3.7& 0.7&3.9\\
Extra resonances &0.4 & 5.9 & 1.9 &2.4 \\
Width Para. &0.6 &0.5 & 1.4  & 4.6\\
Form factor   &0.2 &0.5  &1.3  & 4.6\\
\hline
Total         &0.9 &7.6 & 3.4 & 15.4\\\hline\hline

\end{tabular}
\end{center}
\end{table}

In the measurements of the branching fractions, the systematic uncertainties  from the event selection criteria, the intermediate decays and the number of $J/\psi$ events  are summarized in Table~\ref{sigerr}. 
The systematic uncertainties associated with the MDC tracking and PID efficiencies of kaons are evaluated with the control samples of  $J/\psi\rightarrow K^0_SK^\pm\pi^\mp$. With the clear control sample of 
$\psi(3686)\rightarrow \Xi^-\bar{\Xi}^+$, the uncertainty of the reconstruction of $\Xi^-$, arising from tracking/PID of $\pi^+$, $\Lambda$ reconstruction, the decay-length and mass window criteria, is estimated to be $5.1\%$,
by comparing the MC-data difference. For the yields of signal events, the uncertainties caused by the background shape  and fitting range are estimated by replacing the background shape with a linear order Chebychev polynomial  function
and varying the fitting range. The maximum change of the nominal results is taken as the systematic uncertainty from this source.  
The uncertainty of $\mathcal{B}(\Xi^- \rightarrow \Lambda \pi^-)$ is $0.035\%$, which is negligible, while the uncertainty of $\mathcal{B}(\Lambda \rightarrow p \pi^-)$ is $0.8\%$~\cite{ParticleDataGroup:2024cfk}. 
The  determination of the number of $J/\psi$ events in the data sample  results in an uncertainty of 0.5\%~\cite{BESIII:2021cxx}.

In addition to the above systematic sources, the systematic uncertainties associated with the PWA and their corresponding contributions to the measurements of mass, width and branching fractions are described below.
The input/output uncertainty is evaluated by comparing the input parameters with the corresponding averaged outputs from a series of PWAs with the toy MC samples generated based on the nominal solution.
To assess background-related systematic uncertainties, the alternative PWAs are performed with background levels varied by $\pm$10\% and the maximum change of the results are taken as the systematic uncertainty.
 The uncertainty associated with $\Xi(1820)$ is evaluated by comparing the analyses with and without its inclusion, justified by its minor contribution and spectral location outside the measured mass range.
 The uncertainties related to $\Sigma(1910)$ are quantified by relaxing its mass/width constraints from the PDG values, while the low-significance resonances are sequentially excluded to assess their systematic impacts on the PWA solution. 
The uncertainties from width parametrizations of  $\Xi(1690)$ and $\Xi(1720)$ are evaluated through correlated variations of parameter $f$ from 0.3 to 0.9, while the form factor impacts are studied by varying the $\lambda_1$ and $\lambda_2$ parameters in the range of 1.5 GeV to 3 GeV.  The systematic uncertainty sources and their contributions studied above are all summarized in Table~\ref{sigerr} and Table ~\ref{reserr},  where the total systematic uncertainties are the sum of them added in quadrature by assuming to be independent.

In summary, based on a sample of $10$ billion $J/\psi$ events collected with the BESIII detector, the decay $J/\psi \rightarrow K^- \Sigma^0 \bar{\Xi}^+$ is observed for the first time and  the branching fraction for $J/\psi \rightarrow K^- \Sigma^0 \bar{\Xi}^+$ is determined to be $(2.68 \pm 0.04 \pm 0.17) \times 10^{-5}$.
A detailed PWA is performed to investigate the intermediate excited states contributing to this process.   In case of  the $\Xi(1690)$, the analysis favors a $J^P = \frac{1}{2}^{-}$ assignment, with a mass of  $1695.0\pm0.9\pm0.9$ MeV/$c^2$  and a width of  $12.8\pm1.8\pm 7.6$ MeV, which are consistent with the previous results~\cite{ParticleDataGroup:2024cfk}, but with improved precision, along with a product branching fraction of  $\mathcal{B}(J/\psi \rightarrow \Xi^-(1690) \bar{\Xi}^+  + c.c.)  \times   \mathcal{B}(\Xi^-(1690) \rightarrow K^- \Sigma^0) = (1.86\pm0.54\pm0.68)\times 10^{-5}$. 
Additionally, a new doubly-strange hyperon, denoted as $\Xi(1720)$, is observed with a statistical significance exceeding $10\sigma$. The hypothesis testing among various spin-parity assignments indicates that the $\Xi(1720)$ most likely has $J^P = \frac{3}{2}^+$. Its  mass and width are measured to be  $1721.0 \pm 5.2 \pm 3.4~\mathrm{MeV}/c^2$ and   $31.3 \pm 18.3 \pm 15.4~\mathrm{MeV}$, respectively, and the corresponding product branching fraction
is determined to be $(1.22\pm0.42\pm0.83)\times 10^{-5}$.
Of particular interest is the absence of  any theoretical prediction for such a state around 1.7 GeV/$c^2$ with $J^P =\frac{3}{2}^{+}$. In many theoretical models~\cite{Faustov:2015eba,Menapara:2023rur,Oudichhya:2022off}, the lowest-mass  $\Xi^*$ state with $J^P = \frac{3}{2}^{+}$ is predicted to lie around 1.95 GeV/$c^2$. Similarly, no established state corresponding to $\Xi(1690)$ with  $J^P =\frac{1}{2}^{-}$  has been theoretically anticipated. These observations indicate that our current understanding of $\Xi^*$ resonances remains incomplete, both experimentally and theoretically.
In this case, these results enrich the spectrum of doubly-strange hyperons, provide more information  on probing into their internal structure and  facilitate the understanding of strong interaction dynamics.



The BESIII Collaboration thanks the staff of BEPCII (https://cstr.cn/31109.02.BEPC) and the IHEP computing center for their strong support. This work is supported in part by National Key R\&D Program of China under Contracts Nos. 2025YFA1613900, 2023YFA1606000, 2023YFA1606704; National Natural Science Foundation of China (NSFC) under Contracts Nos. 11935015, 11935016, 11935018, 12075252, 12025502, 12035009, 12035013, 12061131003, 12192260, 12192261, 12192262, 12192263, 12192264, 12192265, 12221005,12225509, 12235017, 12342502, 12361141819, 12475089; the Chinese Academy of Sciences (CAS) Large-Scale Scientific Facility Program; the Strategic Priority Research Program of Chinese Academy of Sciences under Contract No. XDA0480600; CAS under Contract No. YSBR-101; 100 Talents Program of CAS; The Institute of Nuclear and Particle Physics (INPAC) and Shanghai Key Laboratory for Particle Physics and Cosmology; ERC under Contract No. 758462; German Research Foundation DFG under Contract No. FOR5327; Istituto Nazionale di Fisica Nucleare, Italy; Knut and Alice Wallenberg Foundation under Contracts Nos. 2021.0174, 2021.0299, 2023.0315; Ministry of Development of Turkey under Contract No. DPT2006K-120470; National Research Foundation of Korea under Contract No. NRF-2022R1A2C1092335; National Science and Technology fund of Mongolia; Polish National Science Centre under Contract No. 2024/53/B/ST2/00975; STFC (United Kingdom); Swedish Research Council under Contract No. 2019.04595; U. S. Department of Energy under Contract No. DE-FG02-05ER41374.

\bibliography{draftprl-v16}
\end{document}